\documentstyle[psfig]{ioplppt} 
\begin{document} 
\noindent
\title{On the electronic structure of CaCuO$_2$ and SrCuO$_2$}
\author{H.~Rosner\dag , M.~Divi\v{s}\ddag , K.~Koepernik\S ,
S.-L.~Drechsler\dag , and H.~Eschrig\dag}
\address{\dag\ Institut f. Festk\"orper und Werkstofforschung Dresden, 
P.O Box 270016, D-01171 Dresden, Germany}
\address{\ddag\  Charles University, Department of Electron Systems, 
Ke Karlovu 5,  12116 Praha 2, Czech Republic}
\address{\S Max-Planck-Institut f. Chemische Physik fester Stoffe,
D-01187 Dresden, Germany}
%
\begin{abstract} 
Recent electronic structure calculations for the title compounds
performed by Wu et.~{\it al.}~\cite{wu99b} are critically
reconsidered, applying high precision full-potential bandstructure
methods. It is shown that the bandstructure calculations presented by
the authors contain several important inconsistencies, which make
their main conclusions highly questionable.
\end{abstract} 
In a recent paper Wu et.~{\it al.}~\cite{wu99b} presented
bandstructure calculations for the quasi one-dimensional CuO-chain
compound SrCuO$_2$ and the quasi two-dimensional material CaCuO$_2$,
both being of prototypical character and therefore of general
interest.  Wu et.~{\it al.}~used a full-potential linear combination of atomic
orbitals method \cite{wu99a} in the framework of the local spin
density approximation (LSDA) and included on-site Coulomb interaction
corrections (LSDA+$U$).  The authors of Ref.~\cite{wu99b} claim that on the
basis of their full-potential band structure experimental findings can
be well fit with an $U$ of 5 eV, significantly smaller than $U$ values
reported in previous calculations~\cite{anisimov91}.

However, there are obvious inconsistencies and important differences
between the calculations of Ref.~\cite{wu99b} and previous studies
\cite{anisimov91,singh89,mattheiss89,popovic98}, concerning (i) the proper
symmetry in $k$-space, (ii) the widths and the orbital character of
the shown bands, (iii) the total (DOS) as well as the partial
densities of states (PDOS).  Therefore, we reinvestigated the
electronic structures of CaCuO$_2$ and SrCuO$_2$ using two independent, 
well basis converged full-potential bandstructure
methods to find out whether or not the differences mentioned above
could be understood as a consequence of the differences between a
full-potential \cite{wu99b} and the earlier non-full-potential
calculations
\cite{anisimov91,mattheiss89,popovic98}. 
We carried out LSDA bandstructure calculations for CaCuO$_2$ within a
full-potential minimum-basis local-orbital scheme (FPLO)
\cite{koepernik99} and within a full-potential linearized augmented plane wave
(FLAPW) scheme \cite{blaha97}, both in scalar relativistic versions.
(We note that relativistic effects are in the order of 0.1 eV only.)
In the FPLO-scheme, modified Ca 3$d$, 4$s$, 4$p$, (Sr 5$s$, 5$p$
,4$d$), Cu 3$d$, 4$s$, 4$p$, and O 2$s$, 2$p$, 3$d$ states were used
as valence states for CaCuO$_2$ (SrCuO$_2$), the lower lying states
were treated as core states.  The WIEN97-code \cite{blaha97} employs
local orbitals (LO) to relax linearisation errors and to treat the
O-2s and semicore Cu-3p and Ca-3s, 3p states. Well converged basis
sets of over 500 APW functions plus LOs were used.  The radii of the
atomic spheres in the latter case were 1.8 a.u.\ for all atoms.  The
basic calculations were performed with 125 and 90 $k$-points in the
FPLO-scheme and in the WIEN97-code, respectively, for the irreducible
part of the Brillouin zone using the tetrahedron method. We itasize
that the numerical convergence (with respect to the number of
$k$-points $N_k$, the valence basis set, the potential and the density 
representataion) of all calculated properties was carefully
checked.  Following Ref.~\cite{wu99b}, we will discuss first CaCuO$_2$
and afterwards SrCuO$_2$.
 
\underline{CaCuO$_2$}  
First, we will concentrate on the bandstructure and then on the DOS.
Both bandstructures obtained with the FPLO and WIEN97 codes agree
excellently with each other (see Fig.~1(a)) and with previously
published results \cite{singh89, mattheiss89}. 
Considering the bands in Ref.~\cite{wu99b} one
realizes the following points: The authors show a bandstructure with
\textit{orthorhombic} symmetry for the
\textit{tetragonal} crystal structure \cite{vaknin89} (note
the different dispersions in Fig.~1(a) of Ref.~\cite{wu99b} along the
$\Gamma $-(100) and $\Gamma $-(010) direction, respectively, which
must be equivalent for the tetragonal case).  Also, the band
degeneracies at symmetry points are incorrect. As a consequence, the
number of degeneracies in $\Gamma $-(001) direction ($c$-direction)
differs from all other calculations (11 different bands instead of 8
different bands allowed by the crystal symmetry). Contrary to our
results and to the results of Refs.~\cite{singh89, mattheiss89}, the
authors find an additional, third band with sizable dispersion with
its maximum at (001).  We analyzed the orbital character of our bands,
in particular to find out which states are responsible for the
relatively large dispersion in $c$-direction of about 1-2 eV discussed
also by Mattheiss et {\it al.}
\cite{mattheiss89}. Our calculations show that these two strongly
dispersive bands have predominant O 2$p_z$ character with a small
admixture of Cu 3$d_{3z^2-r^2}$ states. In contrast, the
$c$-dispersion of the antibonding band essentially made up by O
2$p$-$\sigma$ and O 2$p$-$\pi$ orbitals with the Cu 3$d_{x^2-y^2}$
orbital is only about 350 meV. Just these states mediate the magnetic
coupling between different layers. This confirms the quasi
two-dimensional character of the magnetic Hamiltonian.

\begin{figure}[t]
\hfill
\begin{minipage}{13cm}
\psfig{figure=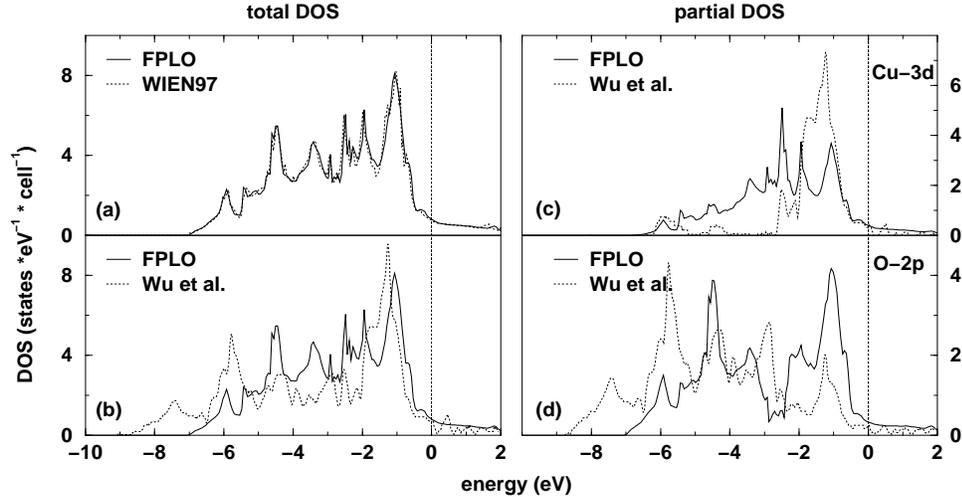,angle=-90,width=13cm} 
\end{minipage}\hfill
\caption{Total DOS (a,b) and partial DOS (c,d) of CaCuO$_2$ calculated
within various bandstructure schemes (see text).}
\end{figure}

In Fig.~1(b) we compare our results for the DOS with those of
Fig.~2(a) in Ref.~\cite{wu99b}. The width of their $pd$-complex is too
large by about 2 eV. The authors attribute this discrepancy to their
choice of an ionic basis, which only means that their calculations are
not basis set converged. The reason for the large discrepancy between
the DOS of Ref.~\cite{wu99b} and our's is evident from Figs.~1(c) and
1(d). Due to the ionic orbital basis used in Ref.~\cite{wu99b}, the O
2$p$ states are shifted downwards by about 2 eV and the hybridisation
with Cu 3$d$ states is consequently reduced.
We attribute simply the reduction of the bandwidth in the LSDA+$U$ in
the calculation of Ref.~\cite{wu99b} to a downwards shift of their Cu
3$d$ states towards the incorrectly positioned oxygen 2$p$ states.
 
\underline{SrCuO$_2$} 
For our paramagnetic calculation the resulting DOS is in excellent
agreement with the DOS reported in Refs.~\cite{popovic98,nagasako97}.
One should note that the occurrence of van-Hove singularities at the
band edges of the antibonding band, due to the nearly one-dimensional
electronic structure of the compound, depends critically on the
sufficiently large $N_k$ used in the calculation.
 
\begin{figure}[t]
\hfill
\begin{minipage}{13cm}
\psfig{figure=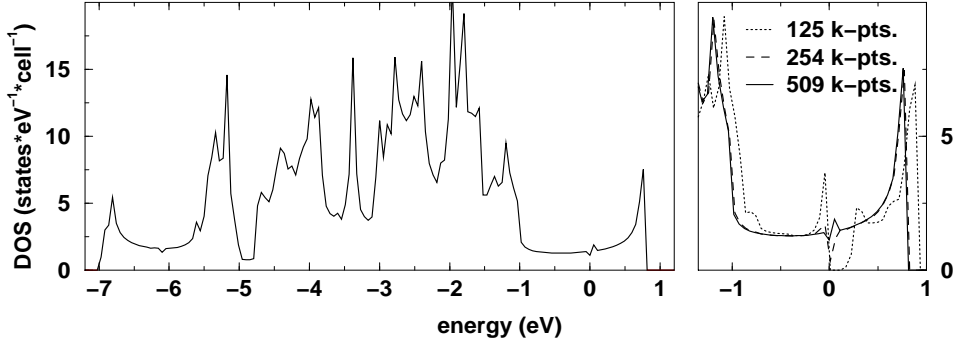,angle=-90,width=13cm} 
\end{minipage}\hfill
\caption{FPLO total DOS for SrCuO$_2$ of the $pd$-complex (left panel) 
and the zoomed region near the Fermi level for different number 
of $k$-points (right panel), calculated in the doubled unit cell.}
\end{figure}

For the supercell calculation, we doubled the unit cell along the
chain direction and started the self-consistent calculation with an
antiferromagnetic arrangement of the Cu spins along the chains
\cite{remark2}. To describe properly some peculiarities related to 
the nearly one-dimensional electronic structure, we made several
calculations varying $N_k$. In particular, we enlarged $N_k$ along the
chain direction. The results are shown in Fig.~2. Due to the nearly
ideal one-dimensional dispersion of SrCuO$_2$, the calculation results
in an erroneous insulating groundstate for an {\it insufficient}
$N_k$. At least for $N_k > 250$, the artificial gap and the related
singularities disappear and the results converge towards those of our
paramagnetic calculation.  Possibly, the gap of 0.55 eV in
Ref.~\cite{wu99b} can be attributed qualitatively to an insufficient
number of $k$-points. The reported relatively big magnetic moment of
0.33 $\mu _B$ is related to this artificial gap and to the unusualy
small hybridisation of the Cu 3$d$ states with the O 2$p$ states.  An
antiferromagnetic solution has been reported also by other authors
\cite{popovic98} though with an even much smaller gap and extremely
small magnetic moments for the SrCuO$_2$-system. Again, we attribute
this gap (notably smaller than that in Ref.~\cite{wu99b} due to a
larger $N_k$) to a still too small number of $k$-points.
\begin{figure}[t]
\hfill
\begin{center}
\begin{minipage}{6cm}
\psfig{figure=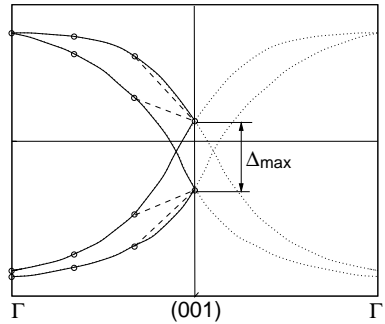,angle=-90,width=5cm} 
\end{minipage}\hfill
\end{center}
\caption{Schematic picture of the folded antibonding bands in $\Gamma$ -
(001) direction in SrCuO$_2$. The solid lines are the correct bands, 
the dashed lines show the interpolated bands for the $k$ points given
by the open circles. For explanation see text.}
\end{figure}
To make this point more clear let us consider schematically the band
structure in the folded zone picture in $\Gamma$ - (001) direction
near the antiferromagnetic Bragg point (see Fig.~3). The weak
hybridisation between the two subchains of the double chain leads to
two slightly split bands. After the folding of these two bands in the
doubled unit cell, new crossing points appear close to the new
symmetry plane.  Therefore, DOS routines result in a wrong
interpolation if the crossing points are not at calculated
$k$-points. The mentioned above splitting $\Delta_{max}$ $\approx$ 150
meV (see Fig.~3) provides an upper bound for the artificial gap
\cite{remark3}.  Hence, for supercell calculations, especially for
quasi one-dimensional electronic structures, one has to be very
careful choosing $N_k$ in applying standard interpolation methods.

In Fig.~4(a) of Ref.~\cite{wu99b} one finds the Cu 3$d$ states in the
PDOS with a too small width as already discussed for CaCuO$_2$.
Moreover, for an orbital projected DOS, it is reasonable to use
symmetry related orbitals with their quantization axis perpendicular
to the CuO$_4$ plaquette. In this representation only the Cu
3$d_{x^2-y^2}$ orbital contributes considerably to the antibonding
band (instead of two Cu 3$d$ orbitals shown in Fig.~4 of
Ref.~\cite{wu99b}), as found for CaCuO$_2$ (Fig.~1(a) of
Ref.~\cite{wu99b}). The LSDA+$U$ procedure used in Ref.~\cite{wu99b}
depends on the basis set representation and would require therefore
the application of the same local orbital symmetry for both compounds
in order to ensure a proper comparison between them.

To summarize, we discussed the main differences between the results of
Wu et {\it al.} and our/or previously published data
\cite{anisimov91,singh89,mattheiss89,popovic98,nagasako97}
basic paramagnetic LDA calculations contain several serious
inconsistencies, all conclusions reported in Ref.~\cite{wu99b} with
respect to LSDA+$U$ are highly questionable. In particular, the
large gap in SrCuO$_2$ at moderate $U$ might result from the
artificial gap found in their LSDA calculations.

\textbf{Acknowledgment:} 
Support for H.R. and M.D.  by the SFB 463 and the Grant Agency of
Czech Republic (grant 202/99/0184 and 202/00/1602) is greatfully
acknowledged.  We thank M.~Richter for stimulating discussions.
 


\begin{thebibliography}{10}
\bibitem{remark2} We suppose that the authors calculated the
same magnetic unit cell and spin structure, but there is no
information about that in Ref.~\cite{wu99b}.
\end{thebibliography}

\end{document}